# Evidence of Ferroelectricity in an Antiferromagnetic Vanadium Trichloride Monolayer

Jinghao Deng[1†], Deping Guo[2,3,4†], Yao Wen[1†], Shuangzan Lu[1,5†], Hui Zhang[1], Zhengbo Cheng[1], Zemin Pan[1], Tao Jian[1], Dongyu Li[1], Hao Wang[1], Yusong Bai[1], Zhilin Li[6], Wei Ji[3,4*], Jun He[1,7*], Chendong Zhang[1*]

[1]School of Physics and Technology, and Key Laboratory of Artificial Micro- and Nano-structures of Ministry of Education, Wuhan University, Wuhan, 430072, China
[2]College of Physics and Electronic Engineering, Center for Computational Sciences, Sichuan Normal University, Chengdu, 610101, China
[3]Beijing Key Laboratory of Optoelectronic Functional Materials & Micro-Nano Devices, School of Physics, Renmin University of China, Beijing 100872, China
[4]Key Laboratory of Quantum State Construction and Manipulation (Ministry of Education), Renmin University of China, Beijing, 100872, China
[5]Hubei Jiufengshan Laboratory, Wuhan, 430074, China
[6]Beijing National Laboratory for Condensed Matter Physics, Institute of Physics, Chinese Academy of Sciences, Beijing 100190, China
[7]Wuhan Institute of Quantum Technology, Wuhan, 430206, China
[†]These authors contributed equally: Jinghao Deng, Deping Guo, Yao Wen, Shuangzan Lu.

*Correspondence author E-mail: wji@ruc.edu.cn (W.J.), He-jun@whu.edu.cn (J.H.), cdzhang@whu.edu.cn (C.D.Z)

**Abstract:** A reduced dimensionality of multiferroic materials is highly desired for device miniaturization, but the coexistence of ferroelectricity and magnetism at the two-dimensional limit is yet to be conclusively demonstrated. Here, we used a $NbSe_2$ substrate to break both the $C_3$ rotational and inversion symmetries in monolayer $VCl_3$ and thus introduced exceptional in-plane ferroelectricity into a two-dimensional magnet. Scanning tunnelling spectroscopy directly visualized ferroelectric domains and manipulated their domain boundaries in monolayer $VCl_3$, where coexisting antiferromagnetic order with canted magnetic moments was verified by vibrating sample magnetometer measurements. Our density functional theory calculations highlight the crucial role that highly directional interfacial Cl–Se interactions play in breaking the symmetries and thus in introducing in-plane ferroelectricity, which was further verified by examining an ML-$VCl_3$/graphene sample. Our work demonstrates an approach to manipulate the ferroelectric states in monolayered magnets through van der Waals interfacial interactions.

Spontaneous polarizations in solid-state materials, including spin and electric polarizations, are crucial characteristics for functional device applications. This scenario has been extended to studies of two-dimensional van der Waals (2D vdW) materials (*1*, *2*). In past years, substantial efforts have been made to validate that either a magnetic (*3*, *4*) or ferroelectric (FE) order (*5*, *6*) can be preserved in the 2D limit. These findings immediately stimulated the interest in searching for vdW monolayers (MLs) that can simultaneously incorporate two (or more) ferroic order parameters, with the goal of achieving multiferroicity (MF) at ML level. This advancement offers prospects for enhanced tunability and flexible integration of polarization configurations with multiple degrees of freedom into complex heterostructures, holding the potential to substantially advance device applications (*2*). Although quite a few candidates were theoretically predicted to host either type-I (*7*, *8*) or type-II (*9*) MF, likely experimental evidence of a multiferroic vdW ML was only revealed for the $NiI_2$ ML based on all-optical methods (*10*). Recent STM studies suggest the persistence of spin-spiral order in the ML $NiI_2$ (*11*), but there remain points of discussion regarding the interpretation of spectroscopic observation related to electric polarization (*12*). It is still an on-going project for the dependable and precise determination of its ferroelectricity (*13–16*). In addition to vdW MLs, the coexistence of ferroelectricity and ferromagnetism was reported for a non-vdW crystal ($Cr_2S_3$) with a single-unit-cell thickness (*17*) during the preparation of this work.

A promising route towards realizing ML MF lies in introducing electric polarizations into known 2D magnets by artificially engineering spatial symmetries. For instance, Huang *et al.* theoretically proposed that Li intercalation or proper charge doping in ML $CrBr_3$ can lead to FE order coexisting with ferromagnetism (*18*). Unlike closed-shell $Cr^{3+}$ ions, open-shell $V^{3+}$ ions enable ML vanadium trihalides ($VX_3$) to exhibit mutable orbital ordering, which, in principle, makes them a more feasible materials family for tailoring lattice symmetries (*19*) and thus inducing electric polarizations (*20*). On the experimental side, bulk forms of $VX_3$ are known to be magnetic with either ferromagnetic ($VI_3$) (*21–23*) or antiferromagnetic ($VCl_3$, $VBr_3$) (*24–26*) configurations. Peculiar orbital orderings (*27*) and the interplay of magnetic and structural degrees of freedom (*28*) have also been found in their bulk crystals. At the ML limit, however, experimental exploration is rare, with sporadic studies reported for $VX_3$ MLs (*28*). Therefore, whether magnetism and electric polarization can be simultaneously retained in ML-$VX_3$ remains unclear.

Here, we epitaxially grow a $VCl_3$ ML on a $NbSe_2$ substrate, in which we report experimental identification of the coexistence of ferroelectricity and antiferromagnetism. Convincing atomic-scale evidence for in-plane (IP) ferroelectricity in the $VCl_3$ monolayer is obtained through scanning tunneling microscopy/spectroscopy (STM/S). This includes key indicators such as severe distortion of the atomic lattice, electric-polarization-induced energy band bending, and tip-induced flipping of the polarization, which collectively verify the presence of ferroelectricity. Moreover, vibrating sample magnetometry (VSM) and density functional theory (DFT) calculations provide consistent evidence that the FE $VCl_3$ ML hosts a bistriped antiferromagnetic (AFM) order ($T_N$ =16 K) with an ***x-z*** easy plane for magnetization. Our DFT calculations further corroborate that directional interfacial interactions (*29*, *30*), specifically Se-Cl covalent-like quasi-bonds, leads to the spontaneous breaking of the both IP $C_3$ rotational and out-of-plane (OOP) inversion symmetries. The broken symmetry induced structural distortion results in an appreciable IP total electric polarization and aids in the selective stabilization of magnetic order. To verify our theory, a control experiment is performed on $VCl_3$/graphene where IP ferroelectricity is absent.—This work experimentally demonstrates a long-sought integration of ferroelectricity into a magnetic monolayer through engineering of van der Waals interfaces.

We started our exploration with molecular beam epitaxy (MBE) growth of $VCl_3$ MLs on freshly cleaved $NbSe_2$ substrates (see Materials and Methods for details). Figure 1A shows a typical STM topography image of an as-grown ML-$VCl_3$ film exhibiting a thickness of approximately 5.6 Å (Fig. S1). A typical tunnelling conductance spectrum indicates a 2.08 eV bandgap for the ML, supporting the semiconducting electronic structures calculated in the band structure calculations (*20*, *31*) with the electronic correlation and orbital ordering considered (for detailed tunneling spectra, see Figs. S2 and S3).In the prototypical crystalline structure of $VCl_3$, V atoms are arranged in a honeycomb lattice, and each V atom is embedded in an octahedron composed of six Cl atoms. Therefore, an STM topography showing triangular protrusions is expected for $VCl_3$, akin to those of epitaxial $CrBr_3$ (*32*) and $CrI_3$ (*33*). Each of these triangles reflects three top-layer Cl atoms surrounding a threefold rotational symmetry center of the V atom. However, our STM imaging of the $VCl_3$ ML shows distinctively different features. Under a relatively large bias, e.g., 1.5 V, the observed protrusions appear as elongated ovals (Fig. 1B). These oval protrusions form a triangular lattice with a periodicity of 5.96 Å, and the primitive vector ***a*** is aligned along the armchair direction of $NbSe_2$ (see Fig. S4). To simplify the latter discussion, we define orthometric ***x-y*** coordinates (as labelled in Fig. 1B) such that ***y*** is parallel to vector ***a*** and ***x*** is defined along the armchair (zigzag) direction of the triangular lattice of oval protrusions ($NbSe_2$). An atomically resolved image shown in Fig. 1C, obtained at a smaller bias of -0.35 V, reveals that one oval protrusion is composed of three top-layer Cl atoms, labelled Cl-1, Cl-2, and Cl-3. Notably, the Cl-2 atom is approximately 0.05 Å higher than the other two Cl atoms in the *z* direction, as illustrated in a close-up 3D perspective plot (upper panel in Fig. 1D). More intriguingly, Fig. 1C indicates pronounced in-plane atomic distortion, where the arrangement of Cl-1 to Cl-3 forms a boomerang-like structure. This configuration is distinctly different from trimerized morphology characterized by an equilateral triangle in Cr-based trihalides (*32*–*34*). In this boomerang arrangement, we define the direction of the protruding Cl-2 atom as the +*y* direction.

Our DFT calculations reproduce this unusual morphology observed in STM. Figure 1E shows a schematic model of a free-standing pristine $VCl_3$ ML, while Fig. 1F illustrates the fully relaxed atomic structure of the most energetically favored stacking configuration of the $VCl_3$ ML grown on $NbSe_2$ (see the detailed discussion in Fig. S6). Upon stacking on $NbSe_2$, the out-of-plane (OOP) inversion symmetry is lifted. In addition, the IP $C_3$ rotational symmetry is also broken owing to interfacial interactions (as elucidated later) that appreciably distort the V-$Cl_6$ octahedra. For clearer visualization, the lattice distortions are slightly exaggerated in Fig. 1D. Such distortion shifts the top-layer Cl atoms off their original positions, allowing us to identify the experimentally defined Cl-1 to Cl-3 atoms in the theoretical model (Fig. 1D and 1F). The middle Cl-2 atom sits higher than the Cl-1 and Cl-3 atoms by 0.02 Å in the calculations (0.05 Å in the experiments), and there is indeed a lateral shift of the Cl-2 atom in the $+y$ direction, forming an IP boomerang-like arrangement. The validity of the theoretical model is further verified by a one-to-one comparison of simulated and experimental constant-$I$ STM images (Fig. 1), while the detailed mechanism for the formation of this distorted structure on $NbSe_2$ will be discussed later. Noted that Fig 1C was obtained following a formerly adopted methodology for scanning ultra-thin insulating/semiconducting layers on metallic substrate, specifically by scanning within the band gap (*35*). This approach helps minimize the impact of spatially varying electronic states on the measurement of morphological heights. Meanwhile, the parameters for comparison in Fig. 1G, which focuses on in-plane morphology, are set at energy levels outside the bandgap to enable theoretical simulations.

Fig. 2A shows a magnified image of the $VCl_3$ flake shown in Fig. 1A, which exhibits a series of parallel boundaries. Two types of domain walls (DWs) were distinguished under a scanning bias of $V_b = 1.5$ V, showing alternating bright (DW-S) and dark (DW-I) DWs. Figs. 2B and 2C show atomically resolved STM images of DW-S and DW-I, respectively. At DW-S, two adjacent Cl boomerangs share a common Cl atom (Fig. 2D), while at DW-I, an isolated Cl atom separates two adjacent Cl boomerangs. The topology of the atomic lattice is continuous across both DW-S and DW-I (see Fig. S7 for details), excluding the possibility of dislocation defects. In addition, the lattice distortions, which are characterized by the $+y$ directions of Cl boomerangs, as labelled in Figs. 2B and 2C, are always mirror symmetric in the adjacent domains, and DW-S and DW-I correspond to the tail-to-tail and head-to-head configurations of the $+y$ coordinates. For simplicity, we used these parallelly aligned DWs in the following discussion, although they are not necessarily parallel, as the lattice distortion is six-fold degenerate in ML-$VCl_3$ (see Figs. S8 and S9 for details).

The alternating bright and dark topographic contrasts of DWs (Fig. 2A) suggest spatial variations in the local density of states across the DWs. Fig. 2E shows a color-coded rendering of the band mapping taken along a pathway across three DWs (white dashed line in Fig. 2A). We used the conduction band, sitting approximately 1.5 eV above the Fermi level, for illustration, which is more evident than the valence band, to show the spatial variation in electronic states (Fig. S2). Clear band bending was observed within each domain. The energy level of the energy band, *i.e.*, the ~1.5 eV conduction band, near DW-S is always higher than that near DW-I. An energy shift of 57 meV was determined by comparing two selected spectra extracted near DW-S and DW-I (Fig. 2F). Noticed that the onset of band edge in Fig. 2F shows slight difference to the tunneling spectrum in Fig. 1A due to the larger tip-sample distance. The reproductivity of the band bending has been verified by multiple different tips across different samples.

Detailed discussion on the influence of different spectroscopic parameters and in the near non-parallel region can be found in Figs. S10 and S11.

This band bending suggests that DW-S and DW-I are most likely negatively and positively charged. Figure 2G shows a schematic summarizing the charge accumulation and band bending at the DWs. Noticed that band bending behaviors between DWs do not show apparent dielectric screening effects (*5*), which can be ascribed to the relatively narrow domains here. In Fig. S12, we show that substantially different band bending behaviors are also observable between opposite edges of the ML-$VCl_3$ flakes (*36*). The estimated screening length in Fig. S12 is around 4-5 nm, which further supports our non-screening scenario since the spacing between DWs are only ranging from 6-10 nm. These lattice-deformation-linked changes in the charge polarities at DWs and/or edges were widely adopted in previous STM studies to recognize IP electric polarizations in 2D layers(*5*, *36*, *37*).

Besides the band bending induced by net electric polarization, another key criteria in STM studies to justify the existence of FE order is the external-field-controlled manipulation of its direction (*5*, *38*). This criterion holds true even for in-plane FE case, such as in studies of SnSe (*39*), where local electric field between the conductive tip and a sample can include an IP component. In Fig. S13, we demonstrate that under specific tunnelling conditions, the IP component of the applied electric field can reversibly switch the direction of lattice distortion in ML-$VCl_3$. This switching induces the controllable movement of domain walls (DWs) and serves as evidence of manipulating local IP electric polarization due to the established correlation between polarization and lattice distortion direction. Based on all above discussions and findings, the conclusion that monolayer $VCl_3$ exhibits in-plane ferroelectricity is supported by a comprehensive and rigorous body of experimental evidence. In terms of the polarization vector $\boldsymbol{P}$, our STS measurements can only reveal the component perpendicular to the DWs, namely, $\boldsymbol{P}_{\perp}$, which was experimentally derived to be ~0.04 $\mu C \cdot cm^{-2}$ (see Methods for details). This polarization strength is comparable to that of known FE vdW layers, such as bilayer $WTe_2$ (~0.19 $\mu C \cdot cm^{-2}$) (*40*) and twisted bilayer graphene/hBN (~0.1 $\mu C \cdot cm^{-2}$) (*41*). Meanwhile, the ferroelectricity arising from structural distortions generally exhibit a high transition temperature. Our current experimental evidence suggests that the transition temperature of $VCl_3$ ferroelectricity is at least above liquid nitrogen temperature and is stable under magnetic field (Fig. S14).

Next, we performed VSM measurements to investigate the magnetic properties of this ferroelectric $VCl_3$ ML (details are discussed in the Methods section and Figs. S14-16). The magnetization-temperature (*M-T*) curves are plotted in Fig. 3A for both zero-field cooling (ZFC) and field cooling (FC) procedures, in which the maximum magnetic field ($\mu_0 H$) reached 1 T. Both *M-T* curves coincide well with each other. An AFM order was characterized by a rapid decrease in magnetization ($M$) with a decrease in temperature ($T$) to below the transition temperature $T_N$. The $T_N$ value was subsequently determined to be $T_N$=16 K by extracting the linear onset of the d$M$/d$T$ curve (Fig. 3B), which is close to the AFM transition temperature of 20 K for bulk $VCl_3$(*24*, *26*). The *M-H* measurements (Fig. 3c) further support the discovered AFM order(*42*, *43*), despite the weak ferromagnetic (FM) features observed near the zero-field region (details are shown in Figs. S15 and S16). In previous studies(*44*, *45*), similar FM features were ascribed to the graphene substrates used for VSM measurements, which were also observed in our measurements on bare substrates with graphene layers (see detailed discussion in Fig. S15). For all our measurements, the magnetization-field curves show similar features for both the IP and OOP magnetic fields, indicating canted magnetic moments in the antiferromagnetic $VCl_3$ MLs. Theoretically, we considered four typical magnetic orders for $NbSe_2$-supported ML-$VCl_3$, as displayed in Fig. 3D, the total energies of which are plotted in Fig. 3e. The bistriped AFM order is at least 2 meV more stable than the other orders. The magnetic moments of the bistriped AFM configuration preferably orient in the ***x-z*** easy plane, exhibiting a negligible (0.005 meV/$VCl_3$) energy difference within the plane (Fig. 3f), while rotating the moments to the ***y***-axis increases the total energy by 0.10 meV/$VCl_3$. The magnetic anisotropy revealed by DFT is well consistent with the VSM results. All these results indicate that ~~BS-~~ bistriped AFM order with an *x-z* easy plane formed in our ML-$VCl_3$. Our theoretical calculations also indicate that the preferred magnetic configuration and the electric polarizations are correlated with the atomic structure deformations (see Fig. S18 for details).

We next discuss the most likely origin(s) of the associated IP ferroelectricity. The preferred BS-AFM configuration results in two types of interfacial Cl atoms, namely, Cl_A and Cl_B1 (Cl_B2), denoted by the green background white letters A and B1 (B2) in the spin-density plot shown in Fig. 4A. Atom Cl_A is sandwiched by two spin-density contours of the same component (yellow isosurface) and lacks spin density in the positive ***y*** direction. However, the two-neighboring spin-density contours of atom Cl_B1 (Cl_B2) are in different spin components (one yellow and one blue isosurface).

This unbalanced spin-density distribution indicates the IP $C_3$ rotational symmetry breaking in the $VCl_3$ overlayer, which induces at least three categories of electron redistribution at the Cl-Se interfaces, as illustrated by the interfacial differential charge density (DCD) at the $VCl_3$-$NbSe_2$ interface (Figs. 4B and 4C). An approximate *y-z* mirror plane could be identified from either the spin density (Fig. 4A) or the DCD (Fig. 4B), where interactions involving Cl_B1 and Se_B1 (orange background white letter B1 hereinafter) are nearly mirror symmetric with those involving Cl_B2 and Se_B2, respectively, e.g., Cl_A-Se_B1 is mirror symmetric with Cl_A-Se_B2. This approximate mirror symmetry requires that the atomic displacements of $VCl_3$ along the *x* direction are nearly mirror symmetric. As a result, the atomic displacements inducing IP electric dipole moments nearly cancel each other under this approximate mirror reflection, as we illustrated quantitively later.

The mirror plane is parallel to the *y*-direction, meaning that the approximate mirror symmetry does not affect electric dipole moments along the *y*-direction. Along path Cl_A-Se_B1 (blue lines in Fig. 4B and 4C), only charge accumulation appears (Fig. 4C). However, directional charge reduction is primarily observed along path Cl_A-Se_A (denoted by the red lines in Figs. 4B and 4C), and alternating charge accumulation and reduction are found along path Cl_B1-Se_B1 (violet dashed line in Fig. 4D). These different charge variations at the Cl-Se interface potentially induce anisotropic atomic displacements for the interfacial Cl atoms and thus likely IP electric polarization in the *y* direction.

In Fig. 4E, we plot the color-coded distribution of the atomic displacements in the *y* direction with respect to the pristine atomic structure of ML-$VCl_3$. The enhanced electron density along path Cl_A-Se_B1 (Fig. 4C) pulls these two atoms away from each other, while the alternating charge accumulation and reduction provide relatively more attractive interactions between Cl_A and Se_A (Cl_B1 and Se_B1). Therefore, the overall result leads to Cl_A undergoing a negative displacement (darkest area in Fig. 4E) and Cl_B1 (Cl_B2) undergoing a slightly positive shift (brighter areas in Fig. 4E) relative to the *y*-axis. As the structures of interfacial and surface Cl atoms are locked by the V-$Cl_6$ octahedron, the surface Cl atoms are shifted upwards for $A_u$ (brightest area in Fig. 4E) and moved slightly downwards for $B1_u$ and $B2_u$ (darker areas) in the *y* direction, which were observed in our STM imaging and DFT simulations.

This nonsymmetric displacement field of cations/anions leads to an ionic contribution of (-0.063, 0.072) $\mu C \cdot cm^{-2}$ to the (*x,y*) vector of the electric polarization. This ionic polarization is well screened, particularly in the *x* direction, well screened by electrons within the ML-$VCl_3$, for which the calculated electronic contribution is (0.060, -0.055) $\mu C \cdot cm^{-2}$. However, the electronic screening is insufficient to fully compensate for the electric polarization along the y direction, leaving a net polarization of 0.017 $\mu C \cdot cm^{-2}$ [(-0.002, 0.017) $\mu C \cdot cm^{-2}$]. Additional screening from interfacial and substrate electrons maintains the electric polarization along the x-direction, keeping its variation within 0.001 $\mu C \cdot cm^{-2}$. In contrast, screening from the substrate further suppresses the polarization along the y-direction, reducing it to 0.014 $\mu C \cdot cm^{-2}$ [(-0.003 0.014) $\mu C \cdot cm^{-2}$], which is consistent with the polarization values and the polarization direction observed in our experiments.

In short, the IP polarization originates from the incomplete electronic screening of the asymmetric atomic displacements that induce polarization along the *y*-direction in the $VCl_3$ overlayer. Such incomplete electronic screening is a well-known feature in 2D materials due to their reduced dimensionality (*46*). These asymmetric atomic displacements are primarily driven by the directional Cl–Se interfacial interactions, often referred to as covalent-like quasi-bonding (*30*, *47*, *48*). This quasi-bonding involves the overlap and hybridization of frontier orbitals between adjacent vdW layers, but differs substantially from true covalent bonds, as the electrons population in the hybridized bonding and antibonding states remain balanced. Thus, despite the enhanced interfacial electronic interactions, the $VCl_3$ monolayer preserves its two-dimensional character on the $NbSe_2$ substrate, as observed in previous studies on other 2D vdW heterojunctions (*49*, *50*). Furthermore, the open-shell configuration of $V^{3+}$, with two *d*-electrons occupying three $t_{2g}$ orbitals, is another essential factor, which contributes to the fragility of the in-plane $C_{3v}$ symmetry. Thus, sufficient directional Cl-Se interfacial interactions can break this symmetry, leading to in-plane structural distortions, favoring a particular orbital ordering and consequently a preferred magnetic order (*51*).

It is of importance to emphasize that this interface-engineered ferroelectricity is distinct from sliding ferroelectricity. In the latter, out-of-plane polarization arises from breaking of inversion symmetry due to interlayer atomic mis-registry, while polarization reversal (*i.e.,* domain wall motion) results from subtle shift in atomic registry due to sliding (*52*). Our case involves lattice distortion-induced electric polarization, which belongs to Type-I multiferroics like $BiFeO_3$ (BFO) and $BiMnO_3$, where ferroelectricity is not directly derived from magnetism. Most Type-I multiferroics exhibit weak magnetoelectric couplings, limiting their applications. However, BFO is an exception allowing electric-field control of magnetic domains through its moderate coupling (*53*). Monolayer $VCl_3$ also has an atomic-structure-mediated appreciable correlation between ferroelectricity and magnetism due to its orbital ordering (Fig. S18).

To verify the crucial role that the directional interaction plays in the IP electric polarization at the $VCl_3$-$NbSe_2$ vdW interface, we theoretically constructed a $VCl_3$ ML on graphene substrate governed by nondirectional interfacial interactions (*54*). Analogous to Figs. 4A to 4D, the spin density, top view of the interfacial DCD and side view of the interfacial DCD are plotted for the $VCl_3$/graphene heterostructure in Figs. 4F to 4H, respectively. Both top views (Figs. 4F and 4G) indicate that the $C_3$ rotational symmetry is nearly maintained for $VCl_3$ on the graphene substrate. While an OOP interface dipole forms at the $VCl_3$-graphene interface (Fig. 4H), the induced IP charge variation is, unlike in the $NbSe_2$ case, nearly isotropic and directionless (Fig. 4F and 4G). Similar isotropic IP interfacial interactions were previously observed at metal–graphene interfaces (*54*, *55*). In the present case, the heterojunction provides an OOP polarization of -1.76 $\mu C \cdot cm^{-2}$ and a nearly zero IP polarization. Noted that such nondirectional interfacial interactions with graphene remain consistent regardless of the specific magnetic configuration in $VCl_3$ (see details in Fig. S19). Controlled experiments were also performed via growth of a $VCl_3$ ML on a graphene/SiC substrate [Methods]. Figure 4i shows a typical STM image of ML-$VCl_3$ grown on graphene/SiC, which indicates that the atomic structure, at least for the surface Cl atoms, is not appreciably distorted and exhibits no polarized domains or well-defined DWs, consistence with our calculated results of the $VCl_3$/graphene heterostructure (refer to Fig. S20 for additional experimental data).

In conclusion, we observed in-plane electric polarizations in a magnetic $VCl_3$ ML selectively grown on a $NbSe_2$ substrate. By combing STM, VSM, and first-principles calculations, we explicitly demonstrate the coexistence IP ferroelectricity and canted bistriped antiferromagnetism. Moreover, we reveal that interfacial Cl-Se interactions in the $VCl_3$-$NbSe_2$ heterojunction breaks both the IP $C_3$ rotational and the OOP inversion symmetries, inducing an IP ferroelectricity. Controlled experiments and calculations for the $VCl_3$-graphene interface verified this mechanism for introducing ferroelectricity, in which vdW interfacial interactions seem to play an elegant role in tailoring the crystalline symmetry and its related emergent quantum phases. Our findings not only provide a potential ML platform for multiferroicity but also refresh the understanding of vdW interface engineering in controlling functional polarizations through multiple degrees of freedom.

## Materials and Methods

**Sample preparations.** The sample growth was carried out within a UHV-MBE system with a base pressure of ~$2.4\times10^{-10}$ Torr. High-quality $VCl_3$ powder (99.5%) was evaporated by a home-built Knudsen cell evaporator. The flux rate was measured to be ~0.15 ML/hour at a growth temperature of 523 K. A fresh $NbSe_2$ surface was acquired by cleaving a bulk $NbSe_2$ crystal in a UHV load-lock chamber at room temperature. During the growth of ML-$VCl_3$ on $NbSe_2$, the substrate temperature was held at 420 K. Postannealing procedures were performed for one hour at 420 K to ensure high-quality crystallization. The growth of ML-$VCl_3$ on graphene and ML-$NbSe_2$ substrates involved growth parameters identical to those for growth on bulk $NbSe_2$. Graphene layers were prepared on 4H-SiC(0001) wafers by a well-established method of vacuum annealing (*56*). Single-layer or bilayer graphene could be selectively formed on the SiC(0001) surface by controlling the annealing time(*56*). Single-layer graphene was used for direct $VCl_3$ deposition (the samples presented in Fig.4d and Fig. S13), and bilayer graphene was used for preparing ML-$NbSe_2$ (*i.e.*, the substrates used in VSM measurements). ML 1H-$NbSe_2$ was grown on graphene/SiC following a previously reported method (*57*). High-quality Nb (99.5%) and Se (99.999%) were evaporated by an electron-beam evaporator and a home-built Knudsen cell evaporator, respectively. High-quality Nb (99.5%) and Se (99.999%) were evaporated by an electron-beam evaporator and a home-built Knudsen cell evaporator, respectively. The flux ratio between Nb and Se was ~1:20, while the substrates were held at 773 K during deposition. To protect the sample from degradation in air, 10 nm Se capping layers were deposited on ML-$VCl_3$ at room substrate temperature before moving it from the ultrahigh vacuum environment to perform VSM measurements.

**STM/STS and VSM measurements.** The STM/STS measurements were carried out on a commercial Unisoku-1300 system. All the measurements were taken at the liquid helium temperature (~ 4.3 K). The PtIr tips used in this work were calibrated by measuring the surface state of a Cu (111) crystal. Tunnelling spectroscopies were performed by utilizing a standard lock-in amplifier with a modulation frequency of 932 Hz. The modulation voltages used are specified in the figure captions. STM images were processed using *Gwyddion* (*58*). Magnetic susceptibility measurements were performed in the Quantum Design PPMS DynaCool with a VSM option. To enhance the signal strength, we stacked 12 pieces of $2\times2$ mm$^2$ ML-$VCl_3$/$NbSe_2$ samples together. ZFC and FC measurements were performed in both in-plane and out-of-plane directions at 1 T, covering a temperature range from 2 to 390 K with increments of 2 K for each step. Magnetization measurements were carried out by sweeping the magnetic field from -9 to 9 T at various temperatures.

**DFT calculations.** Calculations were performed using the generalized gradient approximation in the Perdew–Burke–Ernzerhof (PBE) form (*59*) for the exchange–correlation potential, the projector augmented wave method (*60*), and a plane–wave basis set as implemented in the Vienna ab initio simulation package (VASP) (*61*). Grimme's D3 form vdW correction was considered with the PBE exchange functional (PBE-D3) (*62*) in all structural relaxations. On-site Coulomb interactions of the V $d$ orbitals were considered using a DFT+U method (*63*) with $U$= 3 eV, consistent with the values used in the literature (*20*, *31*, *64*). The structures were fully relaxed until the residual force per atom was less than 0.005 eV/Å. A plane-wave energy cutoff of 450 eV was adopted for the structural relaxation calculation. A k-mesh of 7×13×1 (5×5×1) was used to sample the first Brillouin zone of $VCl_3$ on $NbSe_2$ (graphene). All the vacuum layers adopted (>15 Å) were sufficient to appreciably reduce the image interactions. A $\sqrt{3}\times1$ rectangular lattice ($2\times2$ hexagonal lattice) was adopted for $VCl_3$ on $NbSe_2$ (graphene). Two-layer $NbSe_2$ (graphene) was used to model the substrate, in which the bottom layer was kept fixed and the top layer was allowed to fully relax. Atomic position projections were obtained by subtracting the corresponding atomic positions of bare $VCl_3$ with inversion symmetry from the structure of $VCl_3$ relaxed on $NbSe_2$. Electric polarizations were derived using the Berry phase method (*65*)

**Experimental estimation of the polarization.** By taking the two nearby DWs and the area between them as a parallel-plate capacitor filled with a homogeneous dielectric, the IP polarization could be calculated by the equation $P = (\varepsilon_r - 1)\varepsilon_0 V/d$, where $\varepsilon_r$ is the relative permittivity of $VCl_3$, $\varepsilon_0$ is the electric constant, and $V/d$ is the electric field, in which $V$ and $d$ are the maximum energy shift and the corresponding distance between the two opposite neighboring DWs (*66*). Since the $\varepsilon_0$ of $VCl_3$ is unknown, we took the value for $CrCl_3$ ($\varepsilon_0$ = 8) in our approximation calculations (*67*).

**Acknowledgments:**

We thank Shunping Zhang and Junhao Ge for their discussion during the preparation of revision. Calculations were performed at the Physics Lab of High-Performance Computing and the Public Computing Cloud, Renmin University of China. We thank the Core Facility of Wuhan University and Dr. Hui-Ying Sun for the magnetic measurement.

**Funding:** We thank the supports from the National Key R&D Program of China (Grant No. 2023YFA1406500, 2018YFA0703700), the National Natural Science Foundation of China (Grant No. 11974012 and 12134011), the Strategic Priority Research Program of Chinese Academy of Sciences (No. XDB30000000), the Fundamental Research Funds for the Central Universities, China, and the Research Funds of Renmin University of China [22XNKJ30 (W.J.)]. Calculations were performed at the Physics Lab of High-Performance Computing and the Public Computing Cloud, Renmin University of China. Zhilin Li is grateful for the support from the Youth Innovation Promotion Association of CAS (No. 2021008).

**Author contributions:** J. H.D. and S.Z.L. prepared the samples, and carried out the STM/STS measurements and the data analysis. D.P.G. and W.J. performed the first-principles calculations. Y.W. and H.W. performed the VSM measurements under the supervision of J.H.. Z.B.C, H.Z., Z.M.P., T.J., D.Y.L., and Y.S.B. contributed to the sample preparations. Z.B.C. and H.Z. contributed to the STM measurements. Z.L.L provided high quality single crystals of $NbSe_2$. C.D.Z. initiated the work, advised on the experiments, and provided input on the data analysis. W.J. conceived the theoretical calculations and analysis. J.H.D., D.P.G., Y.W., W.J., and C.D.Z. wrote the manuscript with input from the co-authors.

**Competing interests:** All authors declare they have no competing interests.

**Data and materials availability:** All data needed to evaluate the conclusions in the paper are present in the paper and/or the Supplementary Materials.

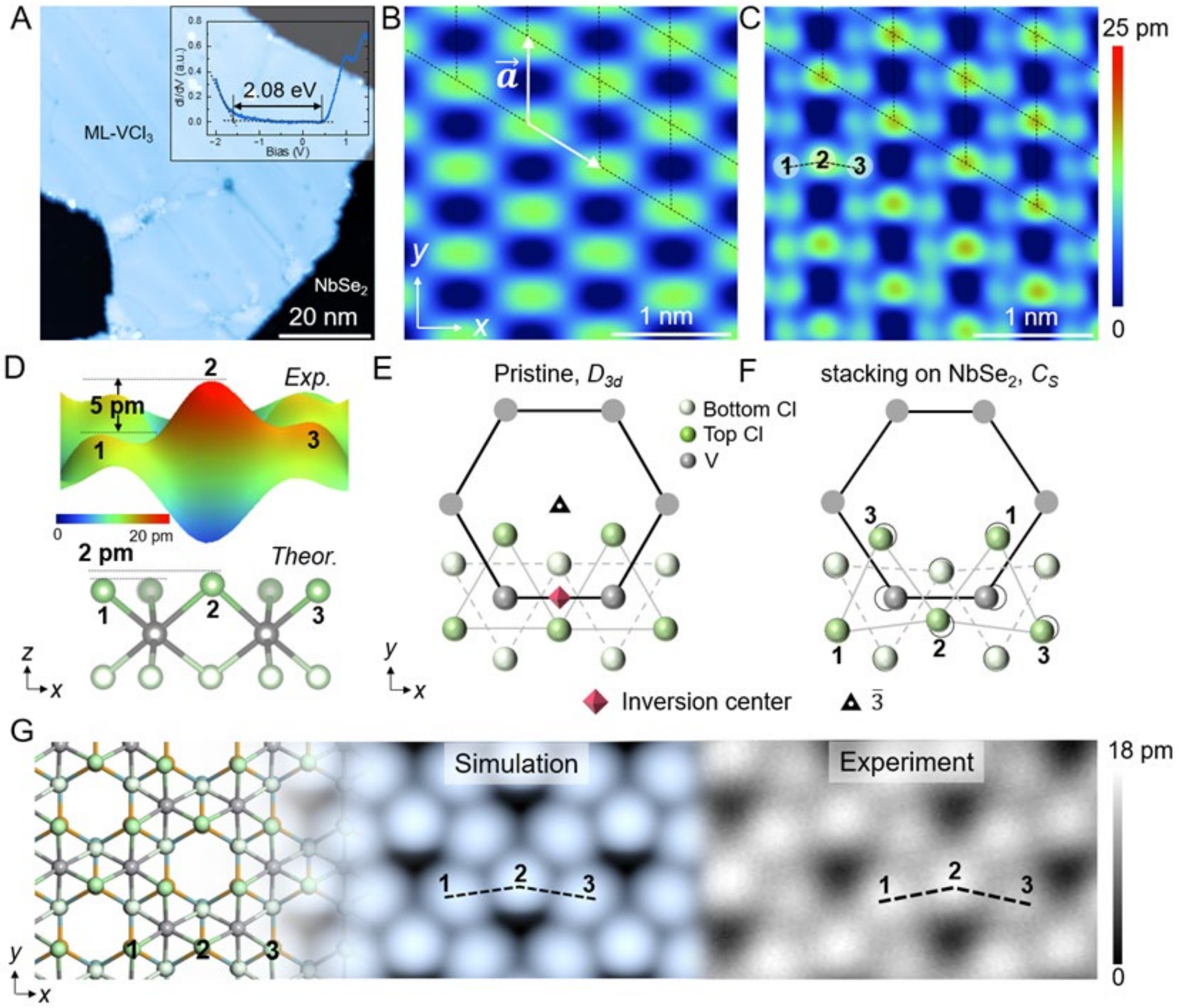

**Fig. 1. Morphology and atomic structure of ML-$VCl_3$ on a $NbSe_2$ substrate. (A)** Large-scale image of an ML-$VCl_3$ flake. Inset: typical d*I*/d*V* spectrum taken on $VCl_3$ (the feedback loop opened at $V_b$ = 1.8 V, $I_t$ = 300 pA and lock-in modulation $V_{rms}$ = 6 mV). The feedback loop opened at $V_b$ = 1.8 V and $I_t$ = 300 pA with lock-in modulation $V_{rms}$ = 6 mV. **(B)** A zoomed-in image of $VCl_3$ and **(C)** is the corresponding atomically resolved image. The dashed black lines indicate the triangular lattice. ***x-y*** coordinates defined with ***x*** along the armchair direction; white arrows represent the lattice constant ***a*.** Three inequivalent Cl atoms are marked by numbers 1, 2, and 3. A black line connects the three Cl atoms in a boomerang-like shape. **(D)** Three-dimensional image zoomed in on a boomerang (upper panel) and a side view of the calculated atomic model (lower panel). The V, top-layer Cl, and bottom-layer Cl atoms are shown in grey, dark green, and bright green, respectively. **(E)** and **(F)** Schematics of pristine $VCl_3$ and $NbSe_2$-supported $VCl_3$, respectively. The $NbSe_2$ lattice is not displayed for simplicity. The red rhombus and black triangle represent the inversion symmetry and threefold rotational symmetry centers, respectively. **(G)** Comparison of the DFT structure (left panel), the simulated STM image (middle panel, simulated under -2.0 V bias), and the experimental STM image. The Nb and Se atoms are shown in blue and orange, respectively. Scanning parameters $V_b$ and $I_t$ are: **(A)** 1.5 V, 20 pA; **(B)** 1.5 V, 50 Pa; **(C)** -0.35 V, $I_t$ = 50 pA, **(G)** -2 V, 50 pA.

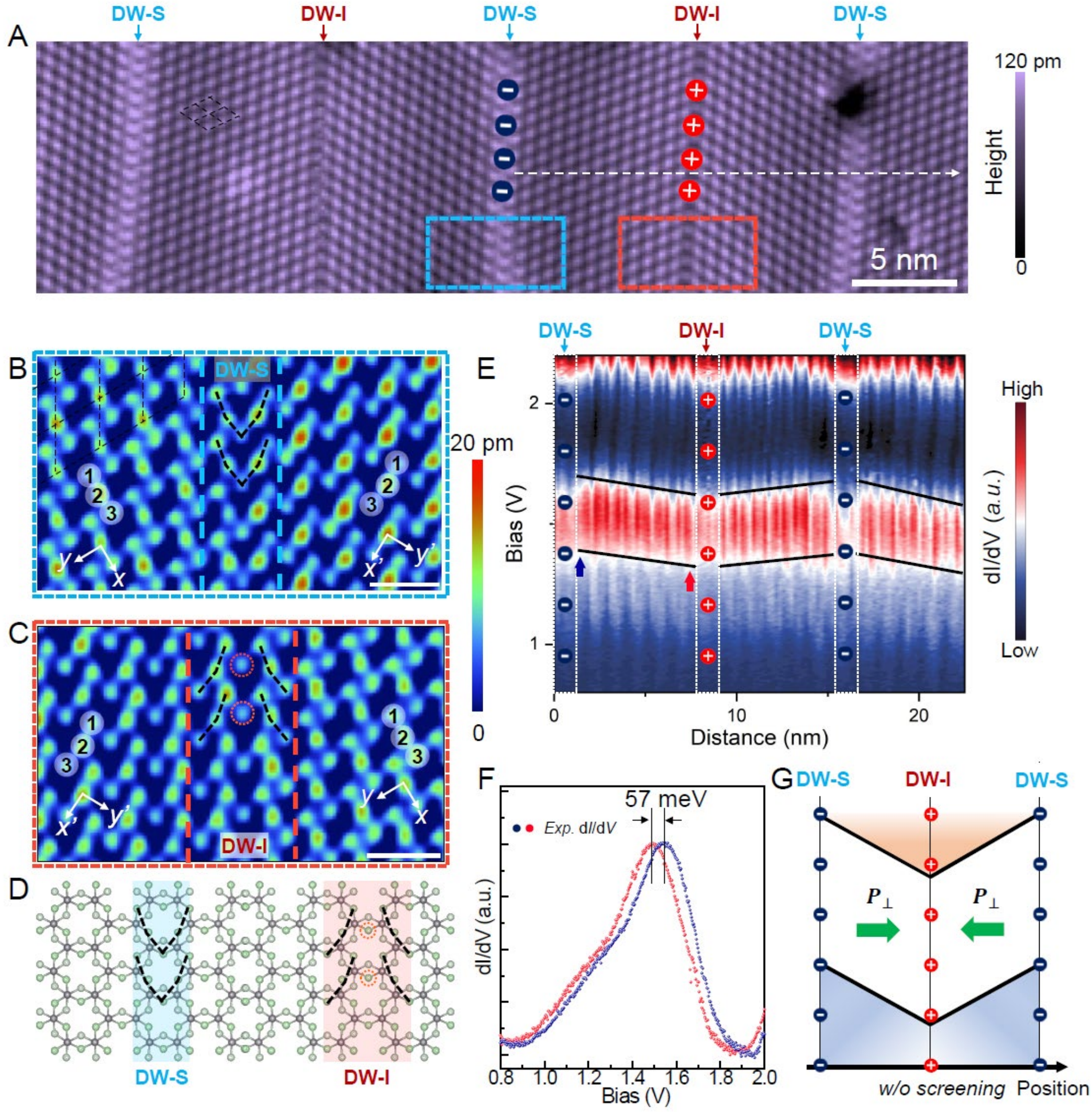

**Fig. 2. In-plane electric polarizations characterized by band bending near DWs. (A)** Image of parallel DWs in ML-$VCl_3$ with DW-I and DW-S marked by "+" and "-" signs for polarization charges. **(B)** and **(C)** Atomically resolved images of DW-S and DW-I, respectively, with dashed grids for unit cells and black boomerang-like lines for the Cl trimers. Red circles in **(C)** highlight the isolated Cl atoms in DW-I. The ***x*** **(*x'*)**-***y*** **(*y'*)** coordinates indicate different orientations of atomic displacements in neighboring domains. Blue and red dashed lines were added as guides to the eyes of the locations of DWs. **(D)** Schematic model of DW-S/I. Cl boomerangs sharing Cl atoms in DW-S and isolated Cl atoms in DW-I are marked. **(E)** Spatially resolved d$I$/d$V$ spectra acquired along the white arrow in **(A)**. The feedback loop opened at $V_b$ = 2.2 V, $I_t$ = 200 pA, and lock-in modulation $V_{rms}$ = 6 mV. The positions of DWs are marked by "+" and "-" markers and white dashed lines. **(F)** Two selected d$I$/d$V$ spectra, corresponding to positions marked in **(E)** and plotted in matching colors. **(G)** Schematic model of the IP electric-polarization-induced band bending without screening. The green arrows and $P_\perp$ denote the components of the polarization vector perpendicular to the DWs. Scanning parameters $V_b$ and $I_t$: (**A**) 1.5 V, 100 pA, (**B**) and (**C**) -0.35 V, 50 pA. Scale bars are 1 nm in (**B**) and (**C**).

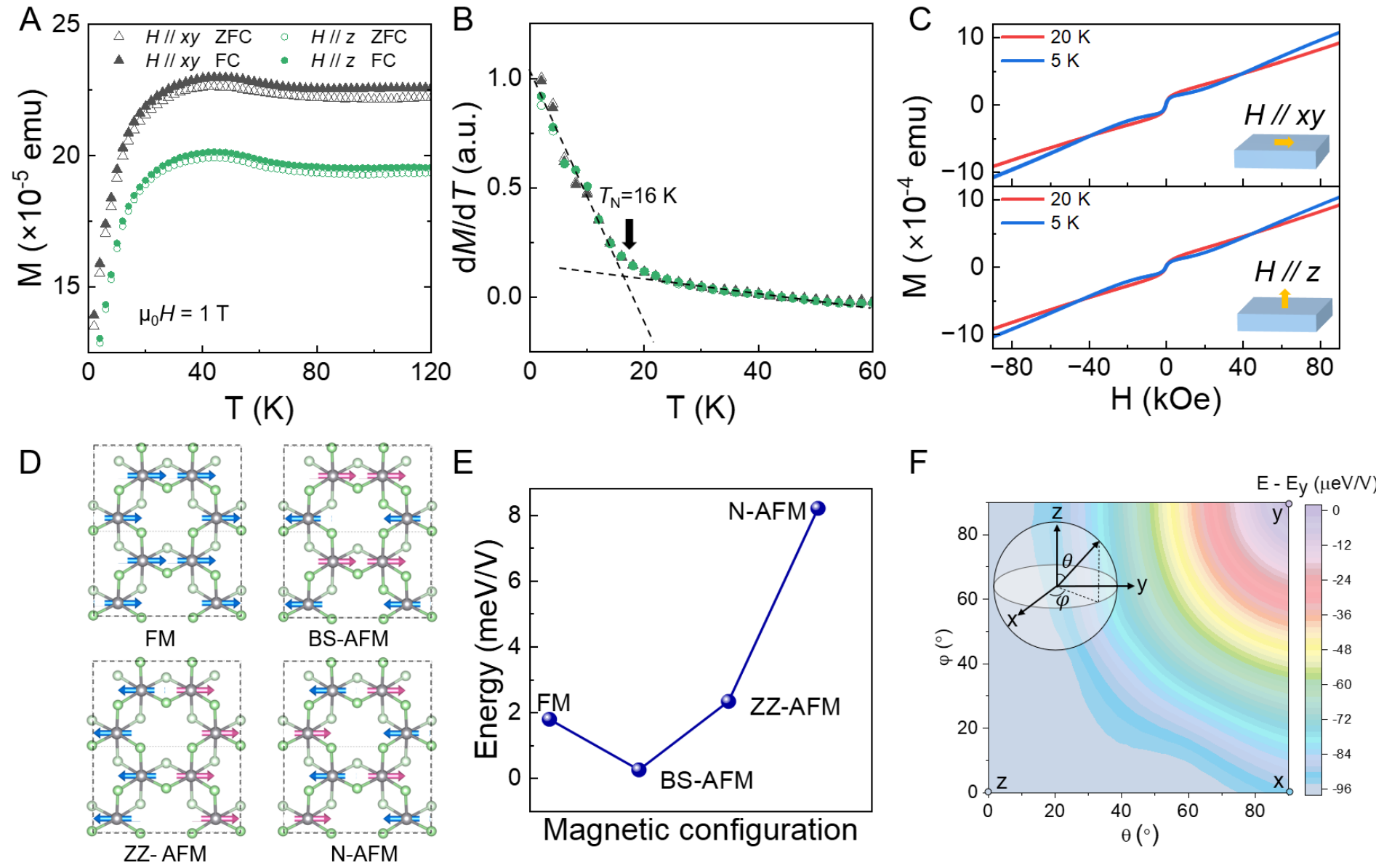

**Fig. 3. Experimental and theoretical investigations of the magnetic order in epitaxy ML-$VCl_3$. (A)** *M-T* curves taken under IP (black) and OOP (green) magnetic fields ($H$ = 1T). ZFC and FC data are shown as hollow and solid dots, respectively. **(B)** d$M$/d$T$ data for determining the AFM transition temperature $T_N$. **(C)** Upper (lower) panels show *M-H* curves taken under IP (OOP) magnetic fields. The red and blue curves correspond to the magnetization curves acquired at temperatures above and below $T_N$ (*i.e.*, 20 K and 5 K, respectively). The details are shown in Fig. S12. **(D)** Illustrations of the ferromagnetic (FM), bistriped (BS), zigzag (ZZ), and Néel (N) AFM configurations of ML-$VCl_3$ on $NbSe_2$. The $NbSe_2$ lattice is not displayed for simplicity. The green and grey balls represent Cl and V atoms, respectively. **(E)** Plot of total energies for the four magnetic configurations. **(F)** Magnetic anisotropy energy mapping of ML-$VCl_3$ on $NbSe_2$ with a bistriped AFM configuration. The coordinate system was defined as shown. The energy of the magnetic moment in the $y$ direction ($E_y$) is set as the zero-point energy.

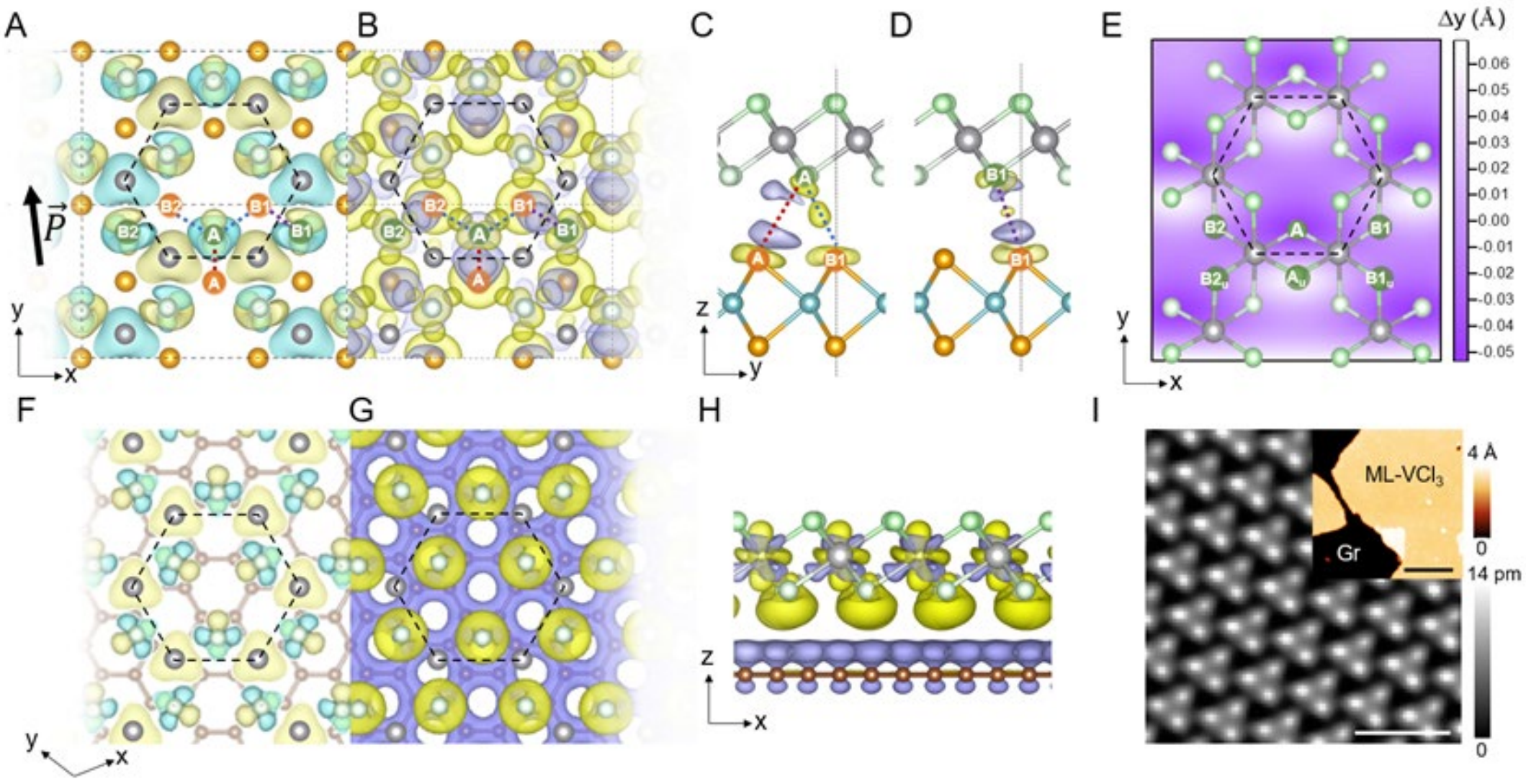

**Fig. 4. Anisotropic charge-transfer-induced IP ferroelectricity and comparison with the $VCl_3$-graphene interface. (A)** Top view of the spin density at the $VCl_3$-$NbSe_2$ interface. Only V atoms, top-layer Se atoms, and bottom-layer Cl atoms are shown. The V, Se, and Cl atoms are shown in grey, orange, and light green, respectively. The yellow (blue) isosurface represents spin up (down), and an isosurface value of $1.2 \times 10^{-4}$ $e$/bohr$^3$ was used. The black arrow in **(A)** indicates the polarization direction $\vec{P}$. **(B)** Top view of the DCD at the $VCl_3$-$NbSe_2$ interface. A yellow (purple) isosurface represents charge accumulation (reduction). The blue dashed lines represent Coulomb repulsion, and the red dashed line represents Coulomb attraction. **(C)** and **(D)** Side views of DCDs at the $VCl_3$-$NbSe_2$ interface solely along path Cl_A-Se_A (red dashed line), Cl_A-Se_B1 (blue dashed line) **(C)** or Cl_B1-Se_B1 (violet dashed line) **(D)**. **(E)** Atomic displacements in the $y$ direction in $NbSe_2$-supported $VCl_3$. The displacements were extracted with respect to the pristine atomic structure. Top (bottom)-layer Cl atoms are marked by $A_u$ (A), $B1_u$ (B1), and $B2_u$ (B2). **(F)** and **(G)** Top view of the spin density and DCD at the $VCl_3$-graphene interface. **(H)** Side view of the DCD at the $VCl_3$-graphene interface. A freestanding bilayer graphene is used in this model with only the top-layer displayed. An isosurface value of $3 \times 10^{-4}$ $e$/bohr$^3$ was used. Carbon atoms are shown in brown. **(I)** Atomically resolved STM image of ML-$VCl_3$ grown on an graphene/SiC substrate ($V_b$ = 0.5 V, $I_t$ = 50 pA, and scale bar: 1 nm). The inset shows a large-area image, where Gr represents the graphene substrate ($V_b$ = 1.5 V, $I_t$ = 20 pA, and scale bar: 20 nm).